\documentclass[twocolumn,showpacs,showkeys,preprintnumbers,amsmath,amssymb]{revtex4}

\usepackage{graphicx}
\usepackage{dcolumn}
\usepackage{bm}
\usepackage{longtable}

\begin{document}

\title{Entropy production, energy loss and currents in adiabatically 
rocked thermal ratchets}
\author{Raishma Krishnan}
\email{raishma@iopb.res.in}
\affiliation{Institute of Physics, Sachivalaya Marg, Bhubaneswar-751005, India}
\author {A. M. Jayannavar}
\email{jayan@iopb.res.in}
\affiliation{Institute of Physics, Sachivalaya Marg, Bhubaneswar-751005, India}

\begin{abstract}
Abstract: We study the nature of currents, input energy and entropy 
production in different types of adiabatically 
rocked ratchets using the method of stochastic energetics. 
The currents exhibit a peak as a function of noise strength. 
We show that there is no underlying resonance or synchronisation 
phenomena in the dynamics of the particle with these current peaks. 
This follows from the analysis 
of energy loss in the medium. We also show that the maxima seen in current 
as well as the total entropy production are not directly correlated. \\
\end{abstract}


 \maketitle
 \section{Introduction}
The subject of noise induced transport has attracted much theoretical 
as well as experimental interest for the past few years 
\cite{julicher,reiman,1amj,special}. The motivation 
for such a study stems from the 
challenge to develop models to explain the reliable unidirectional 
transport observed in biological systems amidst a very noisy environment 
in the absence of bias. Systems that combine the asymmetry 
and nonequilibrium fluctuations to generate systematic motion 
in the absence of a macroscopic bias are termed as ratchets or 
Brownian motors. In thermal equilibrium the principle of detailed 
balance prohibits any net particle current in the system. 
Hence a net current in the absence of any bias can 
appear only as a consequence of the interaction of the particle 
with its noisy  nonequilibrium environment. Thus it is possible 
to extract energy  from the random fluctuations and put it into use. 
These ratchet systems are information engines analogous to the 
 Maxwell's demon which extract work out of bath at the 
expense of an overall increase in entropy 
(or entropy production) \cite{millonas,maxwellamj}. 
There are several ways in which one can incorporate 
nonequilibrium effects arising out of the irreversible 
interaction of the system with its external surroundings. 
This has led to various types of 
ratchets, namely, flashing ratchets, rocking ratchets, 
frictional ratchets, time asymmetric ratchets, etc \cite{reiman}. Extensive 
studies have been carried out on the nature of current and their possible 
 reversals as a function of various physical parameters. These studies are 
found to be useful in identifying proper models for biological motors 
and also to develop machines at the molecular scales including 
nanoparticle separation devices \cite{special}.

The subject of the energetics of Brownian motors or ratchets has 
developed into an entire subfield of its own right \cite{sekimoto,parrondo}. 
A general framework has been developed wherein, the compatibility 
between the Langevin and the Fokker-Planck formalisms used for various types 
of ratchets or motor models and the laws of thermodynamics 
have been proved \cite{parrondo}. 
Using this framework one can readily calculate various physical quantities 
like the efficiency of energy transduction, energy dissipation 
(hysterisis loss), entropy (entropy production), input energy, 
change in internal energy, work etc., in systems far from linear response 
regime into the relam of nonequilibrium domain.  Some recent 
studies have also tried to reveal the relations between two 
completely unrelated phenomena, namely,  stochastic resonance (SR) 
and Brownian ratchets in a formal way through the 
consideration  of  Fokker-Planck equations 
\cite{allison,ref1,ref2,ref3,zhang}. It has been argued that the rate of 
flow of particles in a Brownian ratchet is analogous to the 
rate of flow of information in the case of stochastic resonance. Qian et al. 
have investigated a simple flashing ratchet model for ratchet effect 
as well as SR and have pointed out that the consistency between 
these two phenomena are just due to the existence of circular 
flux in nonequilibrium state \cite{zhang}.

In our present work we have analysed the nature 
of input energy (energy loss) 
and the total entropy production in a class of adiabatically rocked ratchets 
as a function of temperature of the bath (or noise strength). These systems 
exhibit peak in the noise induced current 
(in the absence of any net bias) as a function of temperature. 
The question now arises as to whether this peak is related 
to the underlying resonance due to the synchronization of the position of the 
particle with the external drive induced by the noise. 
Our analysis of input energy $E_{in}$, rules out the presence 
of any resonance features in the dynamics 
of the position of the particle in these systems in the adiabatic regime.

The presence of net currents in the ratchets increases the 
amount of known information about the system than otherwise. This 
extra bit of information comes from the negentropy or the physical 
information supplied by the external nonequilibrium bath. The amount 
of information transferred by the nonequilibrium bath is quantified 
in terms of algorithmic complexity. 
It has been argued that the algorithmic complexity or Kolmogorov 
information entropy is maximum when the current is maximum \cite{san}. 
Since the currents are generated at the expense of entropy we 
naively expect the maxmima in current 
to be related to the maxmima in the overall entropy production 
as a function of noise strength. However, we show that the maxima 
in current and the entropy production do not correlate with each other.

 \section{The Model:}
 We study the motion of an overdamped Brownian particle 
in a potential $V(q)$ subjected to a space dependent medium with 
friction coefficient $\gamma(q)$ and an external periodic force 
field $F(t)$ at temperature $T$. The motion is 
described by the Langevin equation \cite{pareekmcmdan}
 \begin{equation}
 {\dot{q}} = {- {\frac{V^\prime(q)-F(t)}{\gamma(q)}}} - 
{\frac{k_BT \gamma^{\prime}(q)}{2
 {[\gamma(q)]}^2}} + {\sqrt {\frac {k_BT}{\gamma(q)}}} \xi(t)
 \end{equation}
where $\xi(t)$ is a randomly fluctuating Gaussian white noise 
with zero mean and correlation: $<\xi(t)\xi(t^\prime)>\,=
\,2\,\delta(t-t^\prime)$. We take the potential $V(q)$ to be periodic 
in space and is given by   
$V(q)=-sin(q)-(\mu/4)\, sin(2q)$ with the asymmetry parameter $\mu$ 
taking values between $-1$ and $1$. 
Also, we take the friction coefficient $\gamma(q)$ to be periodic: 
$\gamma(q)=\gamma_0(1-\lambda sin(q+\phi))$, where $\phi$ is the 
phase difference with respect to $V(q)$ and the coefficient $\lambda$ 
takes values between $0$ and $1$. The equation of motion is 
equivalently given by the Fokker-Planck equation \cite{riskin}
\begin{eqnarray}
\frac {\partial P(q,t)}{\partial t}&=& \frac {\partial}{\partial q}
\frac{1}{\gamma(q)}\Big[k_BT \frac{\partial P(q,t)}{\partial q}\\ \nonumber
&+& 
[V^\prime(q)-F(t)]P(q,t)\Big]
\end{eqnarray} 

This equation can be solved for the probability current $j$ 
when $F(t)=F_0=constant$, and is given by
\begin{eqnarray}
j&=&\frac{1-exp\,[\frac{-2\pi F_0 }{k_BT}]} 
 {{\int_{0}^{2 \pi}dy I_-(y)}}
 \end{eqnarray}
 where  $I_-(y)$ is given by
 \begin{eqnarray}
 I_-(y)&=&exp\,\left[\frac{-V(y)+ F_0y}{k_BT}\right] \nonumber \\
&\int_{y}^{y+2\pi}dx &\gamma(x) exp\,\left[\frac{V(x)-F_0x}{k_BT}\right]
\end{eqnarray}

In the case of inhomogeneous ratchets (space dependent 
frictional case, $\lambda \neq 0$) \cite{pareekmcmdan,buttiker} 
with the spatial asymmetry parameter $\mu=0$ 
it may be noted that $j(F_0)$ may not be equal to $-j(-F_0)$ for 
$\phi \neq 0, \pi$. This fact leads to the rectification of current 
in the presence of an applied ac field $F(t)$. In these inhomogeneous 
systems directed currents can be obtained even in a  spatially 
periodic symmetric potential. The inversion symmetry in these 
systems being broken dynamically by the space dependent frictional 
coefficient which is periodic in space having a phase lag of $\phi$ 
with the potential profile. In the second case where $\lambda=0$ (purely 
homogeneous case) the net currents are generated due to the 
spatial asymmetry of the potential ($\mu \neq 0$). We 
assume that $F(t)$ changes slowly enough (adiabatic regime), 
i.e., its frequency is smaller than any other 
frequency related to the relaxation rate 
in the problem such that the system is in a steady 
state at each instant of time. For a 
field $F(t)$ of a square wave of amplitude $F_0$, an average current 
over the period of oscillation is given by, $<j> = \frac{1}{2}[j(F_0) 
+ j(-F_0)]$ \cite{pareekmcmdan,kamgawa}. 
In the quasi static limit following the method of 
stochastic energetics it can be shown \cite{kamgawa} that the input energy 
$E_{in}$ (per unit time) is given by 
$E_{in} = \frac{1}{2}F_0 [j(F_0)-j(-F_0)]$.

We also consider another type of ratchet, namely, the time asymmetric 
ratchets \cite{chivalro,phylett,ai} where the driving 
force has zero mean, $<F(t)>\,=\,0$, but is asymmetric in time i.e.,
\begin{eqnarray}
F(t)&=& \frac{1+\epsilon}{1-\epsilon}\, F_0,\,\, (n\tau 
\leq t < n\tau+ \frac{1}{2} \tau (1-\epsilon)), \\ \nonumber
    &=& -F_0,\,\, (n\tau+\frac{1}{2} \tau(1-\epsilon) < t \leq (n+1)\tau).
\end{eqnarray}
The time averaged current in this case is given by
\begin{eqnarray}
<j>=\frac{1}{2}\,(j_1 + j_2)
\end{eqnarray}
with 
\begin{eqnarray}
j_1 &=& (1-\epsilon)\, j(\frac{1+\epsilon}{1-\epsilon}F_0)\\ \nonumber
j_2 &=& (1+\epsilon)\,j(-F_0)
\end{eqnarray}
The input energy $E_{in}$ per unit time for this time 
asymmetric ratchet is given by 
$E_{in}=\frac{1}{2} F_0\,(\frac{1+\epsilon}{1-\epsilon}\,j_1-j_2)$ 
\cite{ai}.  
For the case where $\mu=0,\,\lambda=0$ and $\epsilon \neq 0$ 
the currents are generated in the absence of broken spatial 
symmetry, but in the presence 
of a temporal asymmetric driving with zero mean. This 
type of temporal asymmetry is particularly common in 
biological systems \cite{chivalro}.

 \section{Results and Discussions}
 \begin{figure}[hbp!]
 \begin{center}
\input{epsf}
\hskip15cm \epsfxsize=3in \epsfbox{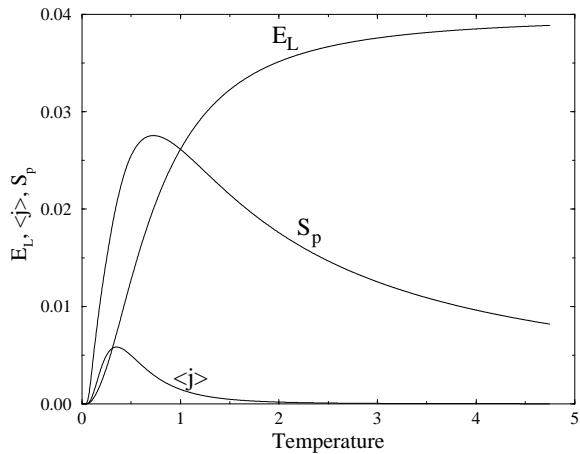}
 \caption{$E_L, <j>$ and $S_p$ vs temperature for $\mu=1.0\,,\,
\lambda=0.0\,,\,\epsilon=0$ 
with fixed $F_0 =0.5$ and $\phi=0.3 \pi$.}
 \end{center}
 \end{figure}
 \begin{figure}[htp!]
 \begin{center}
 \input{epsf}
 \hskip15cm\epsfxsize=3in \epsfbox{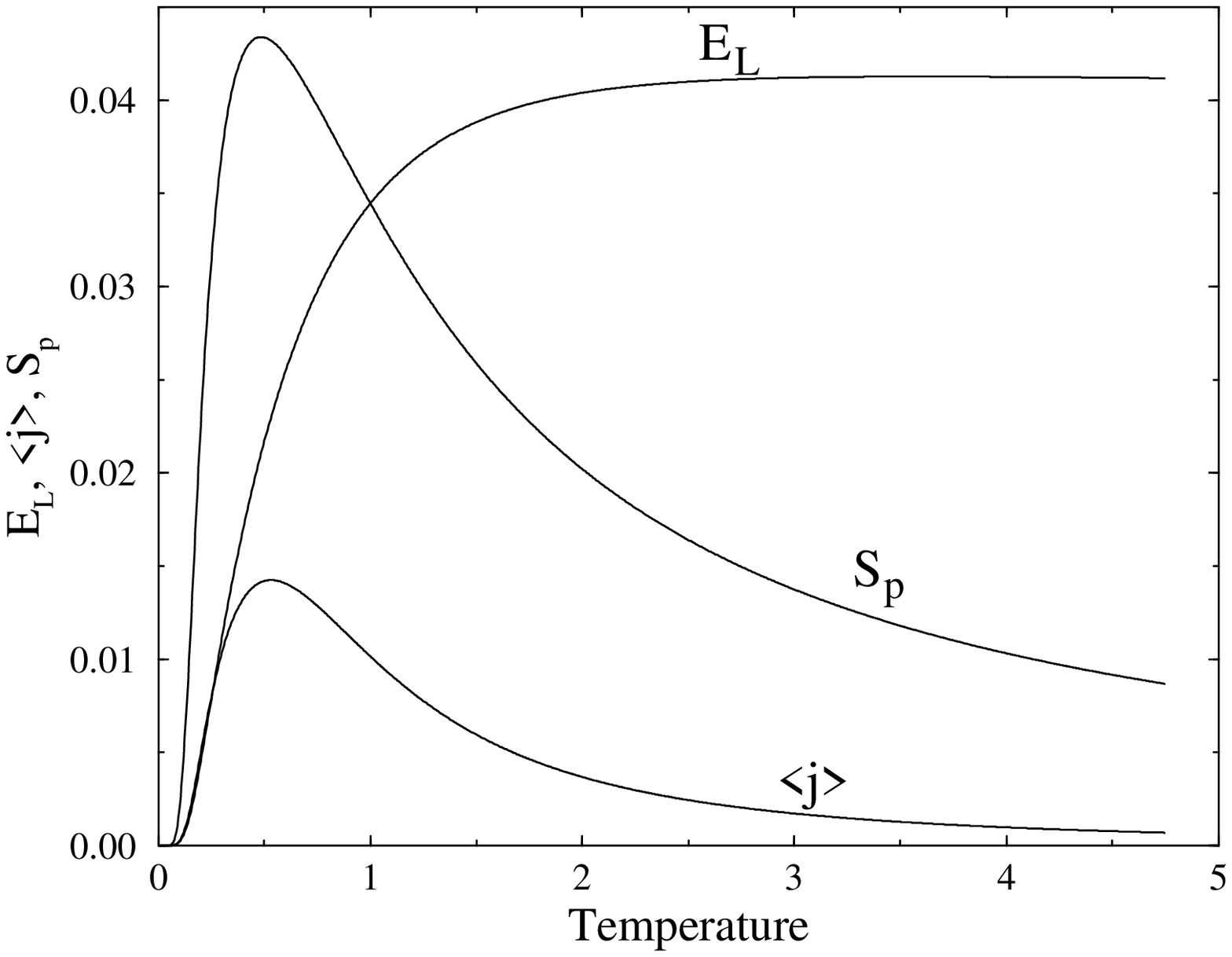}
 \caption{$E_L, <j>$ and $S_p$ vs temperature for $\mu=0\,,\, 
\lambda=0.9\,,\,\epsilon=0$ 
with fixed $F_0 =0.5$ and $\phi=1.3 \pi$.}
 \end{center}
 \end{figure}

\begin{figure}[hbp!]
 \begin{center}
\input{epsf}
\hskip15cm \epsfxsize=3in \epsfbox{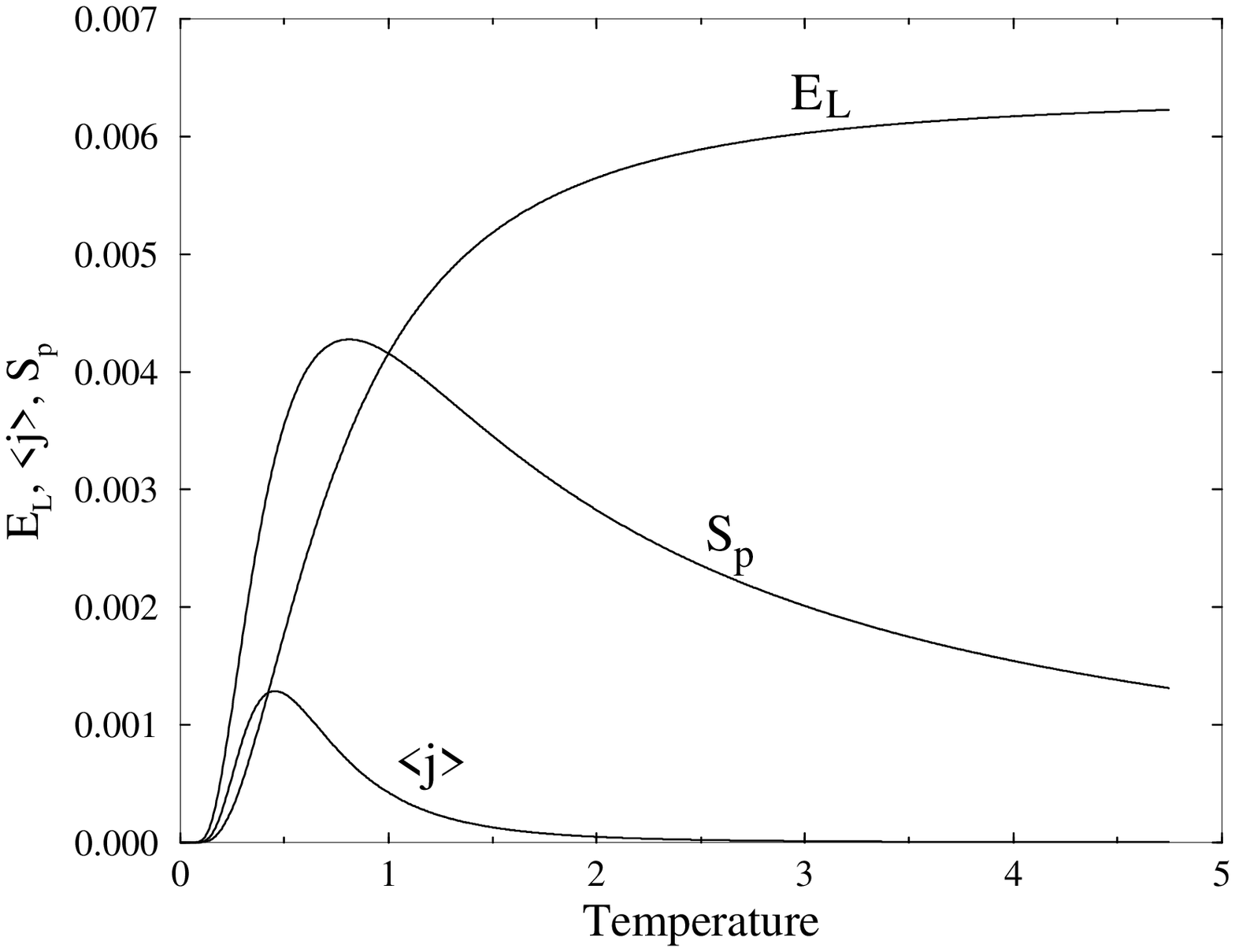}
 \caption{$E_L, <j>$ and $S_p$ vs temperature for $\mu=0\,,\,\lambda=0 
\,,\, \epsilon=0.6$ with fixed $F_0 =0.1$ and $\phi=0.3 \pi$.}
 \end{center}
 \end{figure}

\begin{figure}[hbp!]
 \begin{center}
\input{epsf}
\hskip15cm \epsfxsize=3in \epsfbox{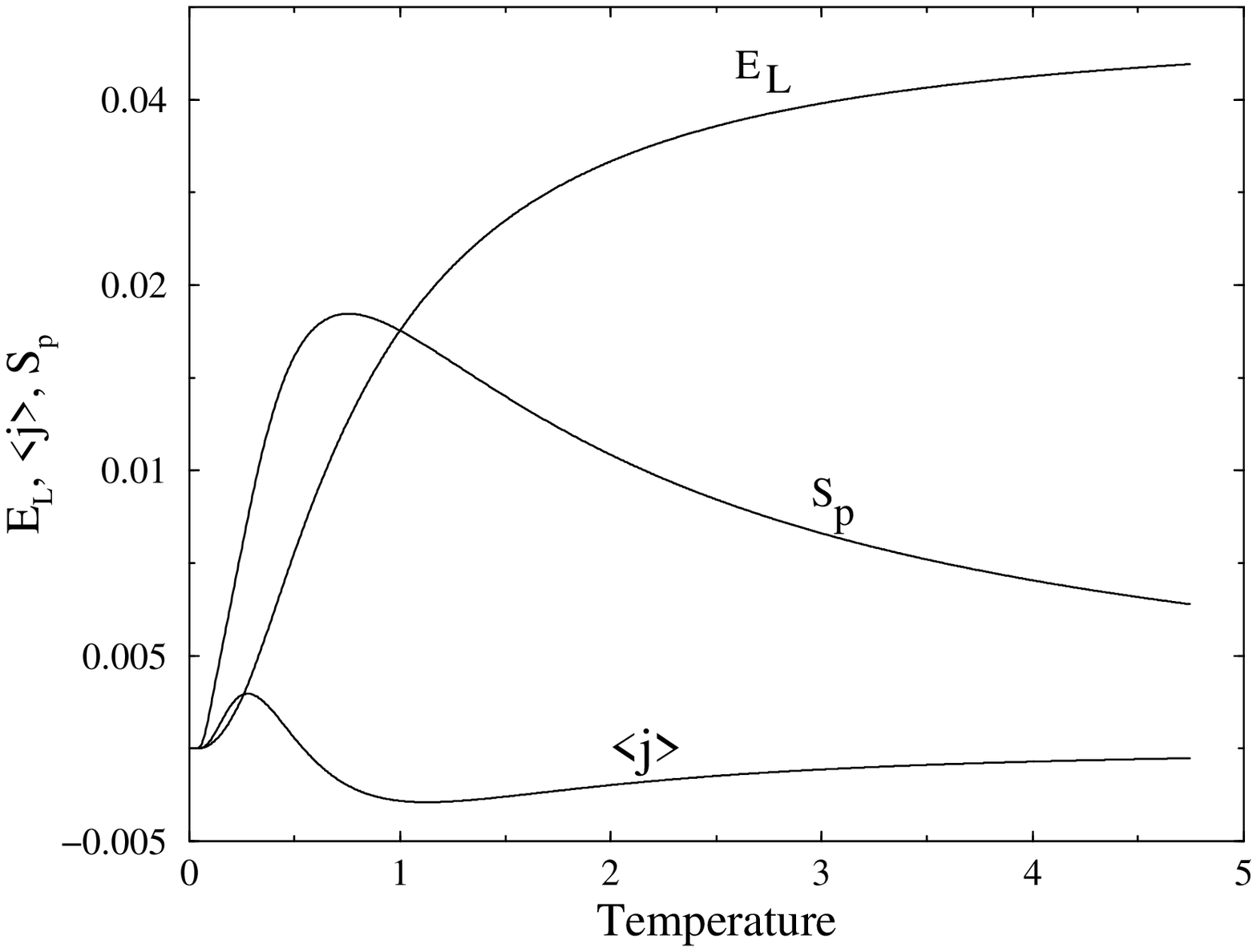}
 \caption{$E_L, <j>$ and $S_p$ vs temperature for $\mu=1.0\,,\,
\lambda=0.9\,,\,\epsilon=0$ with fixed $F_0 =0.5$ and $\phi=0.3 \pi$}.
 \end{center}
 \end{figure}

\begin{figure}[hbp!]
 \begin{center}
\input{epsf}
\hskip15cm \epsfxsize=3in \epsfbox{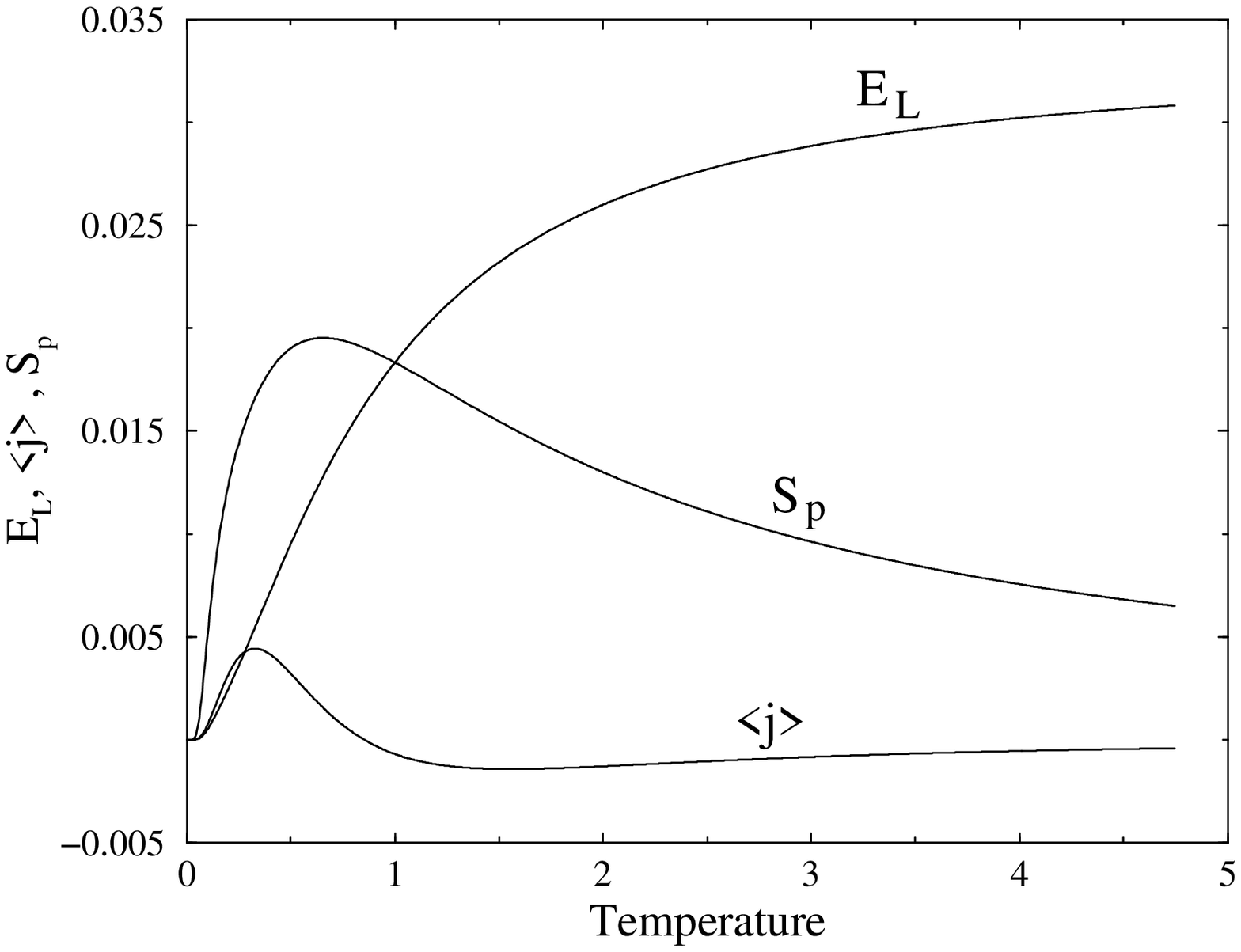}
 \caption{$E_L, <j>$ and $S_p$ vs temperature for $\mu=0\,,\, 
\lambda=0.9 \,,\,\epsilon=0.4$ with fixed $F_0=0.3$ and $\phi=0.3\pi$.}
 \end{center}
 \end{figure}

\begin{figure}[hbp!]
 \begin{center}
\input{epsf}
\hskip15cm \epsfxsize=3in \epsfbox{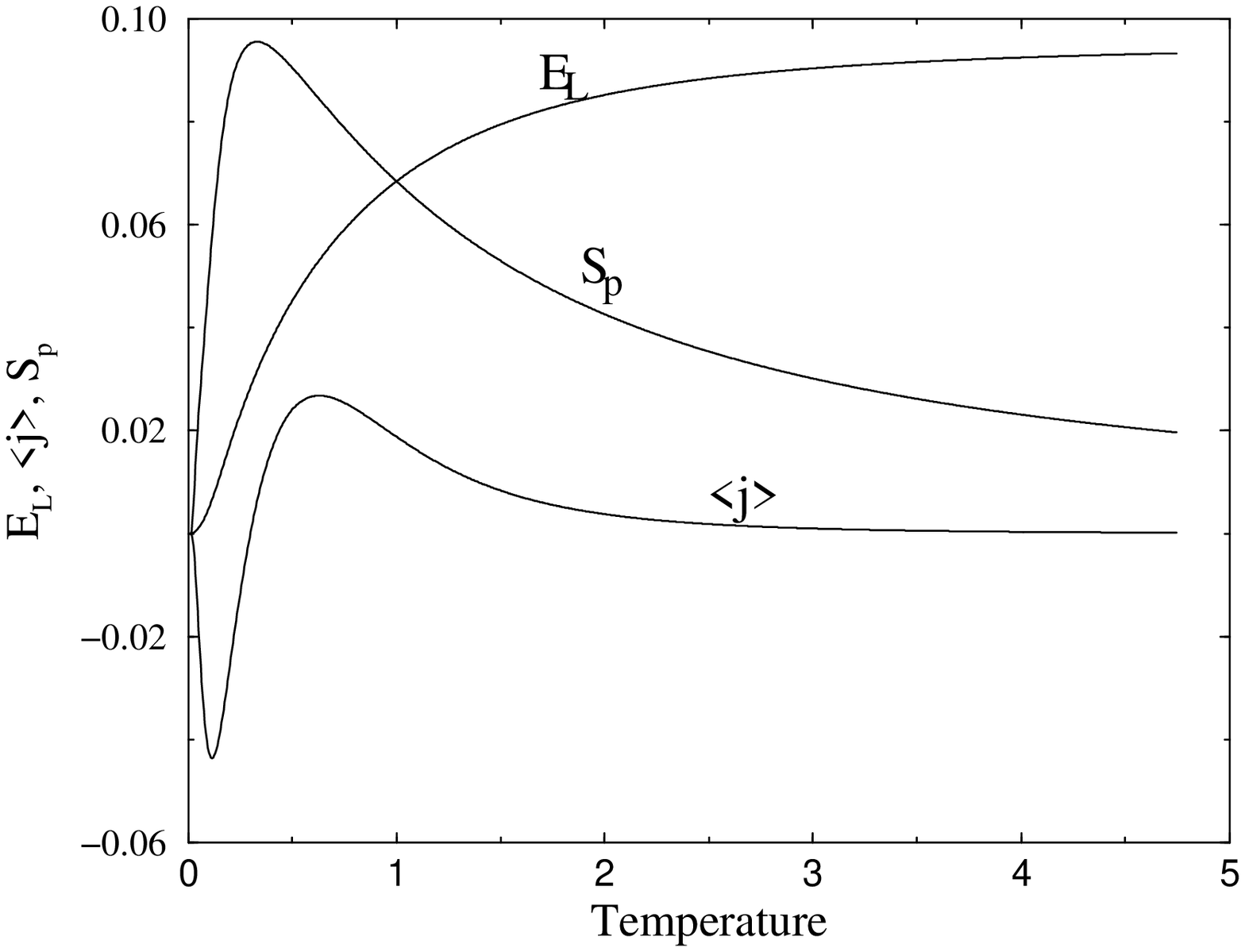}
 \caption{$E_L, <j>$ and $S_p$ vs temperature for $\mu=-1.0\,,\, 
\lambda=0 \,,\,\epsilon=0.25$ with fixed $F_0 =0.6$ and $\phi=0.3 \pi$. 
The current is scaled by a factor of 10 to make it comparable 
with $E_{L}$ and $S_p$.}
 \end{center}
 \end{figure}

\begin{figure}[hbp!]
 \begin{center}
\input{epsf}
\hskip15cm \epsfxsize=3in \epsfbox{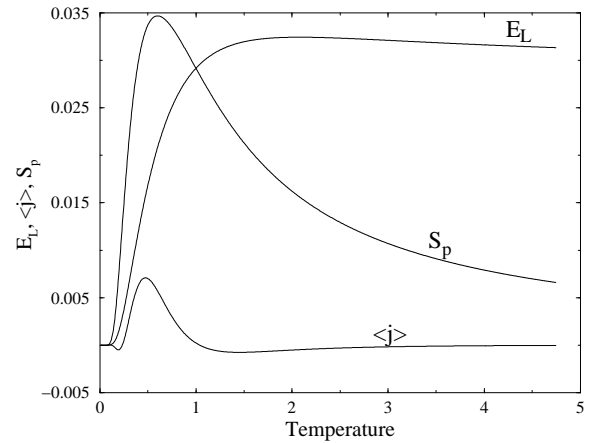}
 \caption{$E_L, <j>$ and $S_p$ vs temperature for $\mu=-1.0\,,\, 
\lambda=0.9 \,,\,\epsilon=0.34$ with fixed $F_0 =0.3$ and $\phi=1.005 \pi$. 
The current is scaled by a factor of 10 to make it comparable 
with $E_{L}$ and $S_p$.}
 \end{center}
 \end{figure}

In the following we analyse all these special classes of adiabatic ratchets 
mentioned above and also their combinations. We study the average 
current $<j>$, the total entropy production $S_p$, 
and the energy loss $E_L$ as a function of temperature 
$T$ (noise strength) for seven different cases of adiabatically rocked 
ratchet systems described below. All these quantities are averaged 
over the period of external drive and are in 
appropriate dimensionless units. It has been argued that for 
the case of a driven double well system input energy is a reliable 
quantity for the identification of SR taking into account 
the detailed comparison between various measures of SR \cite{iwai}. 
Further analysis based on input energy has shown the SR to be a bonafide 
resonance \cite{gammaitoni} in that one obtaines peak 
in the input energy as a function of noise strength 
as well as the frequency of the 
external drive \cite{bonofide}. Thus the peak in the input energy 
represents the matching condition of the escape rate out of the potential well 
(or synchonization in the dynamics of the particle)
and the external driving frequency. It should be noted that as the system 
on the average does not perform any useful work the input energy 
in the steady state equals the energy loss (hysterisis loss) in the medium. 
Hysterisis loss being a good measure to identify SR is 
already known in the literature \cite{hysterisis}. It is quite 
natural that when the system dynamics exhibits a resonance feature 
(or a peak) by tuning certain physical parameters, then at resonance 
the input energy extracted from the source (and the 
concomittant energy loss in the medium) is expected to be high. 
Thus the study of input energy or energy loss is expected 
to reveal the resonances if any in the dynamics of the particle 
as a function of various physical parameters.

In our present work the particle performs a motion in a 
periodic potential in the presence of an adiabatic drive. 
As there is no load applied to the 
system, the system does not perform any useful work or stores energy 
and as a result all the input energy in the steady state will be dissipated 
away. Hence the energy loss in the medium is given by $E_L=E_{in}$. 
$E_L$ inturn is equal to the heat $dQ$ transferred to the bath and 
thus entropy production $S_p=dQ/T=E_L/T$ \cite{parrondo}. 
Thus the total increase in the entropy  (or the entropy production) 
of the bath (universe) integrated over the period of the external drive 
is given by $S_p = \frac{dQ}{T}=\frac{E_{in}}{T}=\frac{E_L}{T}$ 
\cite{parrondo}. As discussed in the introduction, 
currents in the ratchet systems are generated at the expense of entropy 
and thus we expect a correlation between the magnitude of current 
and the total entropy production.

In figures 1 to 7 we have plotted energy loss $E_L$ 
(equal to input energy), average current $<j>$ and entropy production 
$S_p$ as a function of temperature $T$ for various 
types of adiabatically driven ratchet systems. All the physical parameters 
are in dimensionless units and their values 
are mentioned in the figure caption. All these figures 
are representative of the different classes of ratchet 
systems chosen to make our analysis clear.

Figures 1,2 and 3 correspond to the spatially asymmetric case $\mu \neq 0, 
\lambda=0, \epsilon=0 $; inhomogeneous case 
$\lambda \neq 0, \mu=0, \epsilon=0$ and temporal asymmetric case 
$\epsilon \neq 0, \lambda=0, \mu=0$ respectively. In each case currents 
are generated via different types of mechanisms 
arising out of spatial asymmetry in the potential or frictional inhomogenity 
or temporal asymmetry of the external force. Here nonlinearity 
of the system, external drive and asymmetry conspire to 
generate unidirectional currents in the absence of any bias. In all these 
figures (1 to 3) current exhibits a single peak as a function of 
temperature. The input energy being a monotonically 
increasing function of $T$ rules out any resonance in the dynamics of the 
particle as a function of noise strength as discussed earlier. Entropy 
production exhibits a single peak as a function of noise 
strength. Moreover, the peak in the average current, $<j>$ 
and total entropy production $S_p$ does not occur at the 
same $T$. This clearly indicates that maxima in the 
entropy production does not  take place at the same 
value where current is maximum.  To make this point 
explicit in figures 4,5 and 6 we have plotted 
$E_{in}, <j>$ and $ S_p$ as a function of $T$. They correspond 
to a combination of spatial asymmetry and system inhomogenity 
$\mu \neq 0, \lambda \neq 0, \epsilon= 0$; 
system inhomogenity and temporal inhomogenity 
$\lambda \neq 0, \epsilon \neq 0, \mu=0$ and temporal asymmetry 
and spatial asymmetry $\epsilon \neq 0, \mu \neq 0, \lambda=0$ 
respectively. All these figures from 4 to 6 exhibit the 
phenomena of single current reversal or the absolute 
magnitude of current exhibits two peaks. However, the entropy production 
exhibits only a single maxima as a function of noise strength. 
This rules out clearly any correlation between current maxima and the maxima 
in the entropy production. The behaviour of input energy 
rules out any resonance phenomena in these systems as well. 

In fig. 7 we plotted $<j>, E_{in}$ and $S_p$ for a system which incorporates 
frictional inhomogenity, temporal asymmetry as well as spatial asymmetry.
The system with all these combinations exhibit rich variety 
in the nature of current as a function of various physical 
parameters and moreover in some parameter region large 
efficiency of energy transduction is observed \cite{unpub}. For parameters 
given in fig. 7  the current exhibits two reversals as a function of $T$. 
Our observation of double current reversal in the adiabatic 
regime is in itself  a novel phenomena. 
The absolute magnitude of current exhibits three peaks as against to the 
single peak structure in entropy production.  
This again reinforces the fact that maxima in the entropy 
production and current are totally unrelated. 
The behaviour of input energy again rules out the 
resonance dynamics in the system.

In conclusion, by considering different cases of 
adiabatically rocked ratchets we have shown that the 
resonance like feature observed in the nature of current 
as a function of temperature or the noise strength is not related 
to the intrinsic resonance in the dynamics of the particle and 
that the total entropy production does not extremize at the same parameter 
value at which the current exhibits a maximum. Our present 
results are valid only for the case of an adiabatically rocked thermal 
ratchet. This does not rule out the resonance in the dynamics of the 
particle in the nonadiabatic regimes as well as in other 
ratchet systems like flashing ratchets considered 
earlier \cite{allison, ref1,ref2,ref3}. The fact that the noise strength 
at which the current maxima and the maxima in 
entropy production occurs do not coincide may be 
related to the quality of the current. Noise induced currents 
are always associated with a dispersion or diffusion. 
When the diffusion is large then the quality of transport degrades 
and the coherence in the unidirectional motion is lost. The 
coherent transport (optimal transport) refers to the case of 
large mean velocity at fairly small diffusion. It can be 
quantified \cite{sch,dan,low,hanggi} by a dimensionless 
Pe\'{c}let number. For a given 
magnitude of current transport may be coherent or incoherent. Thus analysis 
of the relation between current and the entropy production requires 
not only the magnitude of current but also the quality of 
transport. It is also shown that the algorithmic complexity or the Kolmogorov 
information entropy of the thermal ratchet motion exhibits maxima at the same 
value at which current is maximum \cite{san}. It will be of interest to see 
multiple maxima in algorithmic complexity as a function 
of system parameter in a multiple current reversal regime. 
These studies are expected to reveal a deep connection between 
efficiency, quality of transport, entropy and information. 
Works along these different directions are in progress.

 \end{document}